\documentclass{webofc}

\usepackage[varg]{txfonts}   
\usepackage{hyperref}
\usepackage{url}

\hypersetup{colorlinks=true,citecolor=blue,urlcolor=blue,linkcolor=blue}
\begin{document}
\title{ALICE Event Display}
%
%
\subtitle{from the legacy ROOT-based visualization to the web-based application}

\author{\firstname{Julian} \lastname{Myrcha}\inst{1}\fnsep\thanks{\email{julian.myrcha@cern.ch}} }

\institute{Warsaw University of Technology
          }

\abstract{A Large Ion Collider Experiment (ALICE) is one of the four big CERN experiments at the LHC. The area of interest is the study of the Quark-Gluon Plasma which is produced in heavy-ion collisions. The trajectories of particles created in collisions are reconstructed online and are visualized together with the detector geometry to provide proper augmentation of the presented data. This interactive visualization tool allows 3D visualization of samples taken from the collected data. Starting with LHC Run 3 (from 2022), a newly developed solution has been adopted following the creation of the new ALICE O\textsuperscript{2} Framework. In the first step the data handling part was implemented. The visualization part was developed using technologies from LHC Run 2. This paper presents the process of transition of the visualization component to the modern web based solution. The architecture of the existing ALICE LHC Run 3 online real-time visualization solution is presented. The advantages of the new approach are discussed.
}
\maketitle
\section{Introduction}
\label{intro}

The event visualisation software of the ALICE experiment is used to display the data collected by detectors in a human-understandable way. One of the applications of visualization is the display of images in the control room where it is used, together with other tools, to validate the correctness of the collected data. Starting with LHC Run 3 (from 2022), the ALICE experiment uses the newly developed  O\textsuperscript{2} Framework for online and offline data processing. The development of a new visualization solution resulted from the need to adapt to this framework. Several architecture solutions were taken into account\cite{AtlasBrowser2018}, \cite{GoingStandalone2019}. As a result a transition approach was chosen. To minimize risk of failure, the first version was implemented using the previous ROOT based visualisation approach. However, the architecture was designed for a web based solution. This paper presents how that transition to a web based solution was performed.

\section{Legacy Architecture}
\label{legacy-architecture}

\begin{figure}[h]
\centering
\includegraphics[width=10cm,clip]{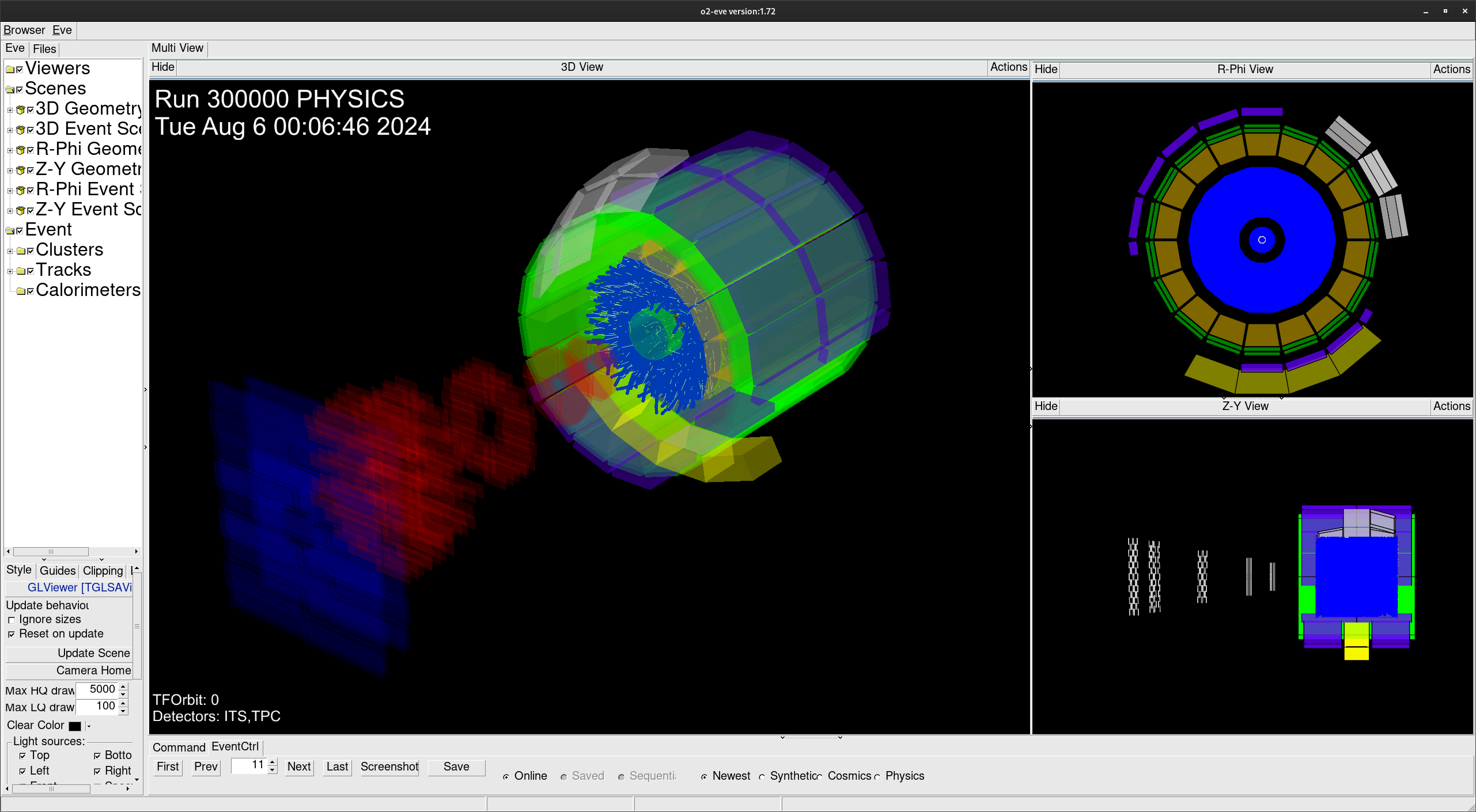}
\caption{Example of a legacy o2-eve visualisation}
\label{fig-o2-eve-legacy}       
\end{figure}

The visualization libraries used at the beginning of LHC Run 3 were chosen as the continuation of the solution used for LHC Run 2 \cite{Niedziela2017}. The visualization tool is written in C++ using the ROOT TEve library (see Fig. \ref{fig-o2-eve-legacy}). The main challenge at this stage of the project was the requirement to adapt to the completely new ALICE O\textsuperscript{2} Framework, which is the source of the data used for visualization \cite{EvolutionOfAlice2019}. As visualization requires data from all detectors it was challenging to write it in parallel with the rest of the code base. 
The software solution was split into two components, with a file-based communication between them. 
The justification for that approach was a clear separation of the components, where development can be done independently and verification can be carried out using unit tests. 

\subsection{o2-eve-workflow - workflow node used to collect data}
\label{o2-eve-workflow}
\begin{figure}[h]
\centering
\includegraphics[width=10cm,clip]{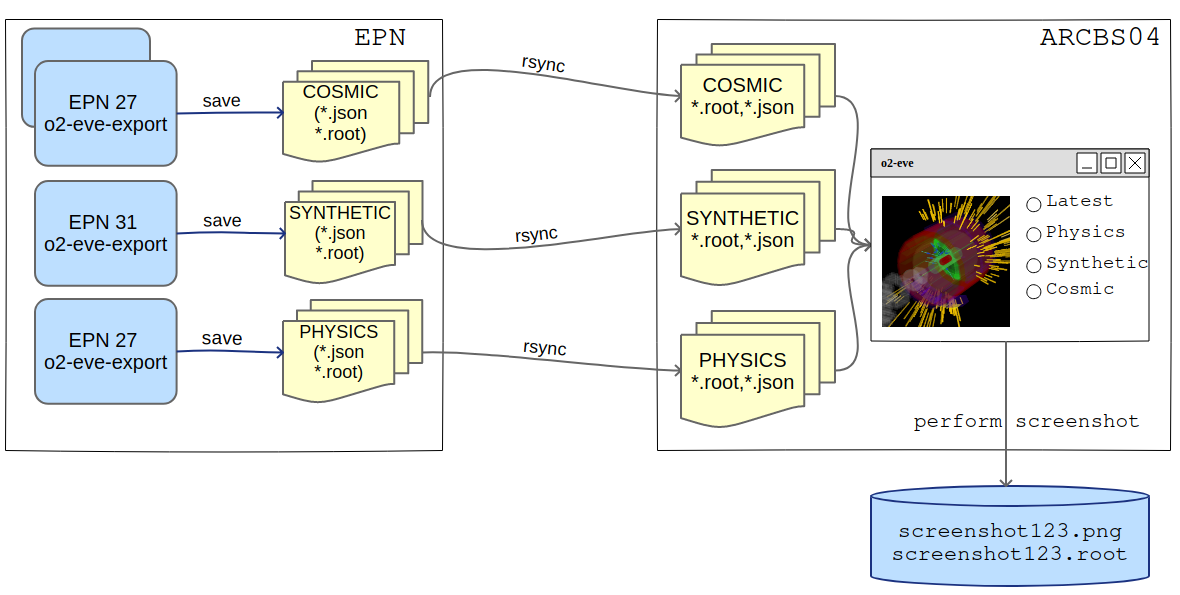}
\caption{Initial visualisation architecture.}
\label{fig-initial-architecture}       
\end{figure}

The purpose of the first component of the solution is to collect the visualization data. This part, written in C++, is one of the ALICE O\textsuperscript{2} workflows  \cite{EvolutionOfAlice2019} which are run on the EPN nodes (see Fig. \ref{fig-initial-architecture}). Depending on the provided parameters, it demands the usage of several workflows which provide the necessary data from each visualized detector. Using all that information, a file is produced and stored on the network file system, containing the following information:
\begin{itemize} 
\item set of general information describing the data to be displayed (example: run number))
\item list of points describing reconstructed tracks, together with physical information about the tracks
\item list of points which describes reconstructed clusters
\item list of data representing information obtained from calorimeters
\end{itemize}

Exact 3D positions of the points for the tracks and clusters are stored. They do not need to be recomputed before visualization, so no physical interpretation is required on the visualization tool to display it. This approach enabled the development of such tools using different technologies.

Files are stored in folders, using the FIFO approach. It provides access not only to the latest data, also to the data collected in a last specified period of time. Separate folders for data taking runs used for different purposes are available:
\begin{description}
\item[PHYSICS] runs that collect collision data 
\item[COSMIC] runs which collect cosmic rays when no collisions are present 
\item[SYNTHETIC] used for testing software configurations
\end{description}

In this way, we have always access to the latest files from each of the run types - the FIFO approach will not override all data of any of the run types.  

For PHYSICS and SYNTHETIC runs, only one EPN node (see Fig. \ref{final-visualisation-architecture}) is used to sample data for visualization. Almost all of the incoming data is skipped - every 2-5 seconds (configurable) a small data segment is saved. A higher rate is unnecessary, as the observer must have enough time to perceive the displayed images. 
For COSMIC runs, all EPN nodes participate, as files from a single EPN node may be too often empty.

\begin{figure}[h]
\centering
\includegraphics[width=10cm,clip]{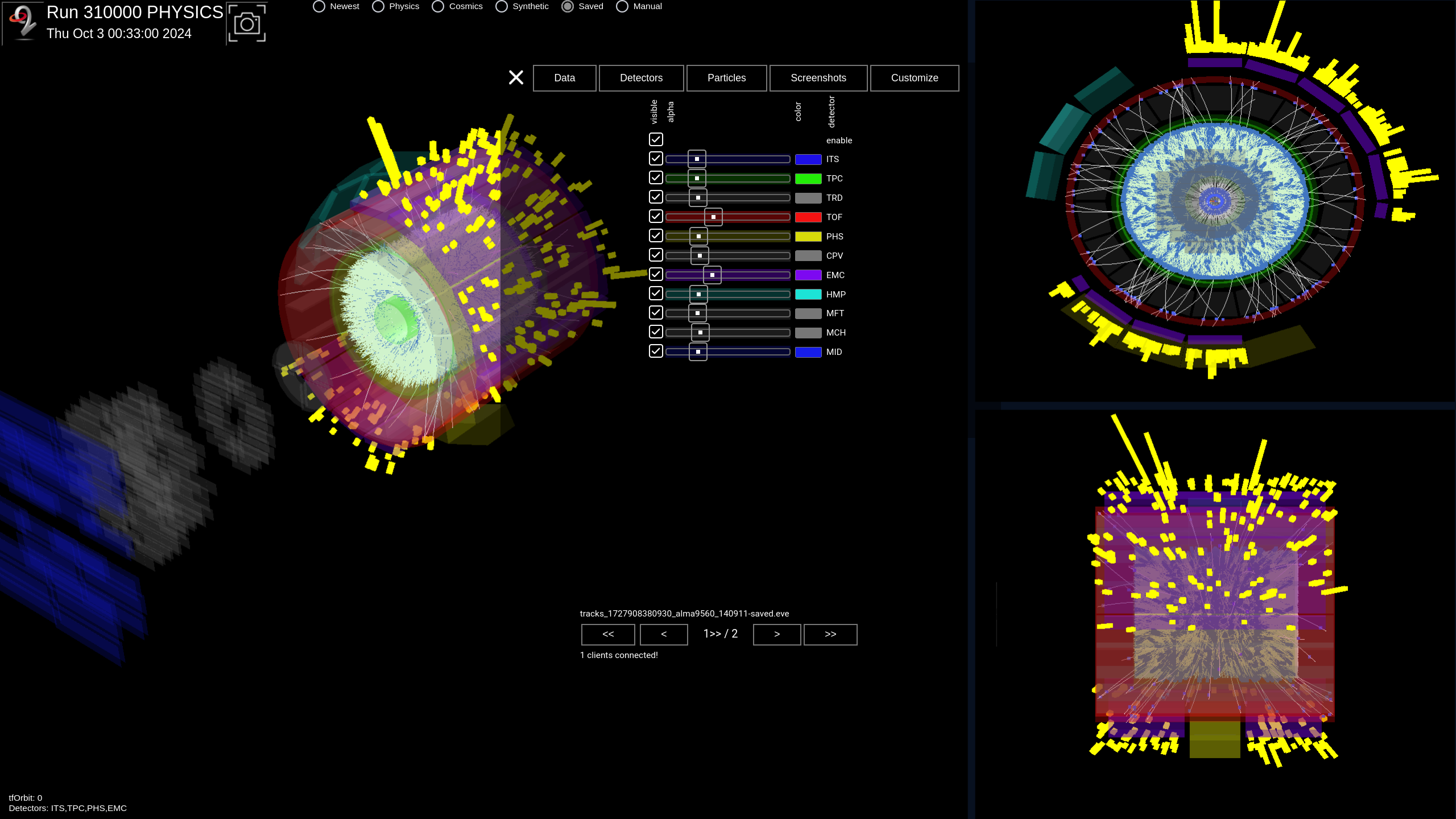}
\caption{Web based visualisation including the parameter setting floating panel.}
\label{final-visualisation-with-displayed-setting-panel}       
\end{figure}

A single file may contain the complete visualization data processed by a given workflow, a so called TimeFrame, which contain data collected during a fixed period of time (currently 2.85~ms). This is convenient for COSMIC runs where there are only a couple of tracks visible during such a period. 
For PHYSICS runs, when for ion-ion collisions there are thousands of tracks in a single collision event, tracks from overlapping collisions become undistinguishable (see Fig. \ref{final-visualisation-with-displayed-setting-panel}). For such runs it is possible to filter out individual collisions from the TimeFrame. They are identified by their primary vertex position. This solves also an additional problem, namely the excessive size of the created files.

\section{o2-eve - visualization in C++ using ROOT TEve}
\label{o2-eve}
In the visualization, tracks, clusters and calorimeter towers are displayed together with a simplified detector geometry which is rendered transparently to not overlap with the data. This provides visual augmentation of the displayed tracks - users can see which detector participated in data collection.
 
The online visualization takes data from files which are mirrored on the visualization computer. The user can select the type of the displayed run or, alternatively, use the option LATEST which combines all three source folders.
The ability to navigate back in the FIFO data allows for extended interaction with a selected visualization, providing up to 10 minutes of playback time, depending on the FIFO settings and the maximum number of files in the folder. The FIFO guarantees that the recorded data are not overwritten immediately by the new ones.

After selecting an interesting visualization, the user can modify the position of the observer, zoom in or out and if they decide that a given data visualisation looks appealing one can create a screen shoot for later usage.
Together with the image file, the file used for creating the screenshot is copied to a folder which stores the screenshots, so if different settings for screenshot are required (background color, resolution, view positions) it may be recreated from the stored file using all other settings.       

\section{File Formats}
\label{file-formats}

During development three file formats were implemented, each serving a different purpose while containing the same set of information.

\begin{figure}[h]
\centering
\includegraphics[width=10cm,clip]{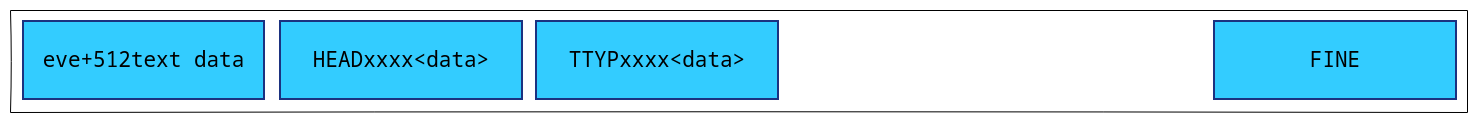}
\caption{Structure of the EVE file format}
\label{format-horizontal}      
\end{figure}

\itemize{ 
\item \textit{JSON} The format was introduced for easy development. Test data may be created in a plain text editor and analysis of the produced data is also simplified. The format is very slow for displaying large files due to its text format which must be properly parsed. For TimeFrame data where the number of tracks reaches the limit set to 50000 tracks, the size of the produced JSON file reaches 800~MB. It was possible to read it using the legacy o2-eve visualization tool but the speed of visualization was largely affected. For the o2-eve-web tool such files are too big due to the JavaScript string size limitation. 
\item \textit{ROOT} Using binary ROOT format resulted in a factor 40 file size reduction compared to the JSON format and very efficient reading by the legacy (ROOT based) visualization tool. For development purposes it was much less convenient but still, using ROOT it is possible to view its contents. For the web based visualization parsing ROOT files adds dependencies on external libraries and also requires their rearrangement before copying into the OpenGL structures.
\item \textit{EVE} This is the final binary format. It currently does not use compression, hence, the file size is up to 2 times larger than the ROOT version of the same data. It was designed in such a way that data can be very efficiently copied into the OpenGL structures with minimal programmatic handling. The structure of the files follows a chunk approach which is used among others in PNG file format (see Fig. \ref{format-horizontal}). This makes it expandable.  It is possible to create private chunks and when not recognized, they will be ignored in visualization. It is also possible to create filters converting one type of the chunks into another - for example adding compression can be done by adding new chunk type and converting into this chunk reading and writing software. The first chunk has fixed size and is a free description of the file (totally ignored by the reading procedures). Using the Linux command line <head> tool (or similar approach on other systems) it can be displayed on the console. So, it is possible to read basic information about a file content without any proprietary software on any computer.
}

\begin{figure}[h]
\centering
\includegraphics[width=5cm,clip]{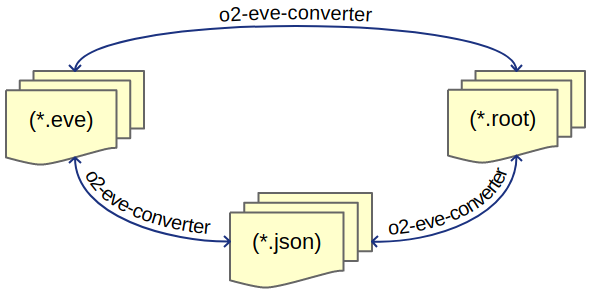}
\caption{The o2-eve-converter can convert between all three supported formats.}
\label{o2-eve-convert}       
\end{figure}

A o2-eve-converter command line tool has been implemented which can convert, without any information loss, between all of the above file formats (see Fig. \ref{o2-eve-convert}).
Using the converter it is still possible to profit from the advantages of the JSON format during the development phase.

\section{Final Visualisation - o2-eve-web}
\label{final-architecture}

The main goal of the visualization architecture design was its ability to perform a smooth replacement of the ROOT based solution with the new one. This happened in 2024 and for a couple of months both solutions were used interchangeably. After adding handling of the EVE format to the legacy visualization it can now be used on the new data format.

\begin{figure}[h]
\centering
\includegraphics[width=10cm,clip]{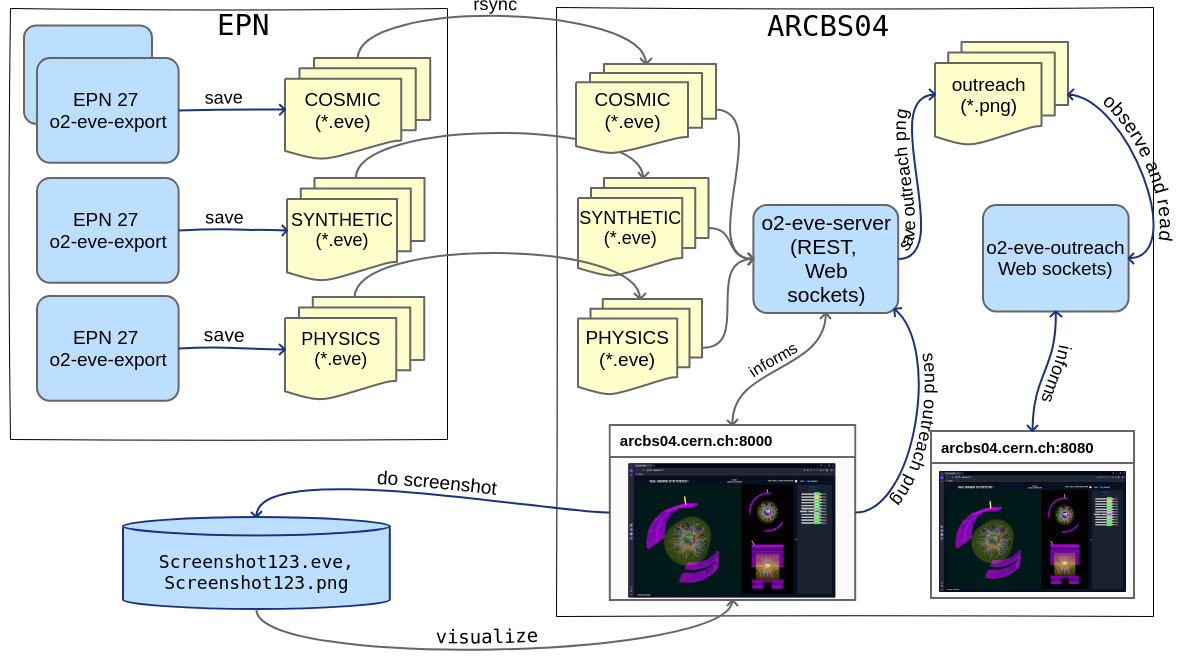}
\caption{The final visualisation architecture.}
\label{final-visualisation-architecture}       
\end{figure}

In the o2-eve-web visualisation tool, the web server observes the data folders and notifies the browser using web sockets technology each time when the content of the folders changes (see Fig. \ref{final-visualisation-architecture}).

The web visualization was implemented using the THREE.js library. This is an industry standard used by thousands of programmers all around the world. Hence, newly identified problems are promptly fixed. 
The functionality of the library is well suited for on-line visualization in ALICE. The limitations of the library, described in Ref. \cite{RenderCore2024} is not affecting the presented solution. On the other hand, the size of the community supporting the library is considered a big advantage.

In the OpenGL world, replacing the graphical model every frame is not a typical situation.
The required speed of visualization was achieved by proper caching and reuse of already allocated OpenGL structures, which eliminated the need to reallocate them for each event.  

\begin{figure}[h]
\centering
\includegraphics[width=12cm,clip]{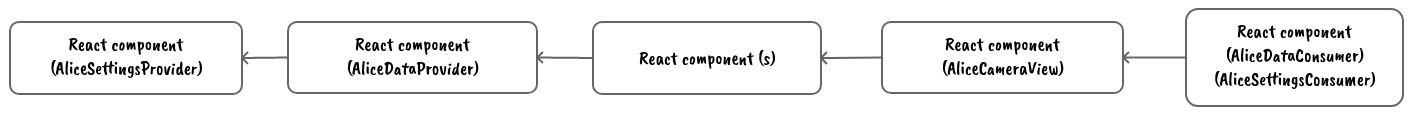}
\caption{Simplified Providers/Consumers in o2-eve-web visualisation}
\label{fig-context-horizontal}       
\end{figure}

The internal architecture was created using another very popular library - REACT.js, which provides a component based architecture. The visualisation component was split into several provider/consumer pairs responsible for various functionalities (see Fig. \ref{fig-context-horizontal}), which can be easily reconfigured or replaced. 

From the development point of view, the web visualisation technology has plenty of advantages. Hot reloading makes development much faster, as there is rarely the need to restart the application after applying changes. Splitting the solution into provider/consumer pairs made it possible to develop them independently from other parts of the application. It is possible to test if the provider produces correct information using separate test projects. The same can be done for the consumer by using mock providers which produce well defined test information.

For web visualization, the web server operates in the ALICE Control Room. Each user of the web visualization has their own copy of the program in their browser, so only data transmission can have a performance impact. 

The client which is run in the ALICE Control Room is treated in a special way. It has additional configuration settings available (like path to the folder used by the outreach web server discussed in point \ref{outreach-web-server}). It also produces screenshots for that server into this folder.

The client is also informed about every change in the source folders on the web server, so it sees all new data. To reduce the possible impact of remote clients (which could cause server overload) other clients are informed with a frequency that decreases with the number of connected clients. This approach ensures that appropriate server performance is maintained in the Control Room at the expense of data refresh rate on other client computers. If this were a problem, the architecture provides an immediate solution - files should be synchronized on an additional server. Server in Control Room would serve only one client and additional server would serve all external clients in a way that would not affect the performance of visualization in the Control Room.

Configuration parameters (colors, resolution of the screenshots) are stored in the browser local storage and is preserved between application runs. Screenshots made by the user (together with the display data - the approach is the same as for ROOT based visualization) are stored locally on the user machine. That display data may be later loaded manually to recreate screenshots.

\subsection{Outreach web server}
\label{outreach-web-server}
Using interactive web visualization sometimes has non-obvious inconveniences. This happens when interactivity is not needed, like for various exhibition purposes, where the requirement to provide setup parameters is a problem, not an advantage. For such a scenario a second web server is serving screenshots created by the first one (see Fig. \ref{final-visualisation-architecture}). This also guarantees, that screens connected from exhibitions show exactly the same data as presented in the ALICE Control Room.

\subsection{Performance}
\label{performance}
The implemented solution is capable to display up to 50 000 tracks and the same amount of clusters in much less than a second. After loading the data the user can interactively change the point-of-view position and the zoom level scale and this works smoothly. 

One issue  occurs when the user connects remotely. EVE files have sizes from 3MB (typically) up to 40 MB (for full TimeFrame data for PHYSICS run) and the network transmission time may have an impact.

\section{Conclusions}
\label{conclusions}
Event visualization in the ALICE experiment is used online in the Control Room. Starting from 2024 a new web application is used. The transition from the previous ROOT-based approach was seamless, thanks to careful planning and well-thought-out strategic decisions made at the outset of development. The new web-based architecture achieved the same performance as the previous ROOT based approach. We achieved the flexibility to connect to the same data from additional machines (from outside of the Control Room) without the need to install any software - only a web browser is needed. 

From the maintenance and future development point of view, the new architecture is much more convenient. The time needed to apply changes was greatly reduced compared to the C++ ROOT based solution.  
  
Foreseen future developments include additional visualization components as possible replacement of the existing one with new visual effects such as path animation or detector model cross-sections. Thanks to the component-based architecture of the present solution, such extensions should not affect the rest of the visualization application.
The EVE file format can also be extended with additional blocks, like those for performing data compression functions. This requires research to see if the gain resulting from transmitting less data will not be lost by the additional time required for decompression.

\bibliography{CHEP-2024-Alice-Event-Display}
\end{document}